\documentstyle[]{amsart}
\newcommand{\R}{{\Bbb R}}
\newcommand{\Z}{{\Bbb Z}}
\newcommand{\Nb}{{\Bbb N}}

\newcommand{\A}{{\cal A}}
\newcommand{\M}{{\cal M}}
\newcommand{\T}{{\cal T}}
\newcommand{\Hh}{{\cal H}}

\newcommand{\N}{{\cal N}}

\newcommand{\E}{{\cal E}}
\newcommand{\Mg}{(M,g)}

\newcommand{\HS}{\operatorname{HS}}

\newcommand{\SB}{\operatorname{SB}}

\newcommand{\End}{\operatorname{End}}

\newcommand{\Aut}{\operatorname{Aut}}
\newcommand{\Tr}{\operatorname{Tr}}
\newcommand{\tr}{\operatorname{tr}}
\newcommand{\rank}{\operatorname{rank}}

\newcommand{\weak}{\operatorname{weak}}
\newcommand{\range}{\operatorname{range}}

\newcommand{\expit}{\exp\operatorname{it}}

\newcommand{\spec}{\operatorname{spec}}

\newcommand{\Id}{\operatorname{ Id}}

\newcommand{\weaklim}{\operatornamewithlimits{weak^*-\lim}}

\newcommand{\sa}{\sigma_A}
\newcommand{\sd}{\sqrt{\Delta}}
\newcommand{\pis}{\Pi_\Sigma}

\newsymbol\upharpoonright 1318
\newsymbol\dhr 1317

\theoremstyle{plain}

\newtheorem{thone}{Theorem 1}[section]

\newtheorem{thtwo}{Theorem 2}[section]
   
\newtheorem{ththree}{Theorem 3}[section]

\theoremstyle{definition}

\theoremstyle{remark}

\title{Quantum ergodicity of $C^*$ dynamical systems}
\author{Steven Zelditch*}
\date{September 1994}
\thanks{ *Partially supported by NSF grant \#DMS-9404637.}
\address{Johns Hopkins University, Baltimore, Maryland  21218}

\begin{document}
\maketitle

\addtolength{\baselineskip}{1pt} 

\setcounter{section}{-1}
\section{Introduction}

The purpose of this paper is to generalize some basic notions and results on quantum 
 ergodicity  ( [Sn], [CV], [Su],  [Z.1], [Z.2]) to a wider class of $C^*$ dynamical
systems $(\ A, G, \alpha)$ which we call  {\it quantized Gelfand-Segal systems}
 (Definition 1.1).  The key feature of such a system is an invariant state $\omega$ which
in a certain sense is the barycenter of the normal invariant states.
By the Gelfand-Segal construction, it   induces a new system $(\A_{\omega},
G, \alpha_{\omega}),$ which will play the role of the classical limit.  Our main abstract result
(Theorem 1 ) shows that if $(\A, G, \alpha)$ is a quantized GS system, if the classial limit
is abelian (or if $(\A, \omega)$ is a ``G-abelian" pair), and if 
 $\omega $ is an ergodic state, then  ``almost all" the ergodic normal invariant states
 $\rho_j$ of the system tend to $\omega $ as the ``energy" $E(\rho_j)\rightarrow
\infty$.  This leads to an intrinsic notion of the quantum ergodicity of a quantized GS system
in terms of operator time and space averages
  (Definition 0.1),  and to the result that a quantized GS system is quantum 
ergodic if its classical limit is an ergodic abelian system
(or if $(\A, \omega)$ is an ergodic G-abelian pair) (Theorem 2).  
  Concrete applications   include a simplified proof
of quantum ergodicity of the wave group of a compact Riemannian
manifold with ergodic geodesic flow, as well as extensions to manifolds
with concave boundary and ergodic billiards,  to quotient Hamiltonian systems on
symplectic quotients and to ergodic Hamiltonian subsystems on sympletic subcones.
More elaborate applications will appear in forthcoming articles: to manifolds with
general piecewise smooth boundary and ergodic billiards in [Z.Zw] , and in [Z.5]
to quantized ergodic contact (or contactible)
transformations acting on powers of a line bundle (including quantized
hyperbolic toral automorphisms acting on spaces of theta functions).

To state our results, we will need to introduce some terminology and
notation.  We will also briefly review some relevant background on
quantum ergodicity and on $C^*$ dynamical systems, with the aim of
clarifying the connections between the two.

Quantum ergodicity, in the sense of this paper, is the study of quantum
dynamical systems whose underlying classical dynamical systems are
ergodic.  For instance, the wave group $U_t = \expit \sqrt\Delta$ of a
compact Riemannian manifold $\Mg$ is the quantization of the geodesic
flow $G^t$ on $S^*M$.  The basic problem is to determine the asymptotic
properties of various invariants of the spectrum $\{\lambda _j\}$ and
eigenfunctions $\{\phi_j\}$ in the limit $\lambda _j\rightarrow \infty$,
under the condition that $G^t$ acts ergodically with respect to the
normalized Liouville measure $d\mu$ on $S^*M$.  For some of the many
heuristic and numerical results we refer to the recent survey of Sarnak
[Sa].

From the $C^*$ algebra point of view, a quantum dynamical system is a
$C^*$ dynamical system $(\A, G, \alpha)$ where $\A$ is a  $C^*$-algebra, and 
$\alpha :G\rightarrow \Aut(\A)$ is a representation of $G$ by
automorphisms of $\A$.  We will always assume $\A$ is unital and
separable, that $G$ is amenable and that the system is covariantly
represented on a Hilbert space $\Hh$.  That is, we will assume there is a
representation $\pi:  \A\rightarrow {\cal L} (\Hh)$ of $\A$ as bounded
operators on $\Hh$, and a unitary representation $U:G\rightarrow U(\Hh)$
such that $\alpha _g(A) = U^*_g\pi(A) U_g$. Representations are understood
to be continuous.  Henceforth we will denote
$\pi(A)$ simply by $A$.  For terminology regarding $C^*$ algebras we
follow [B.R] and [R].

As is evident, the notion of quantum ergodicity which we intend to generalize is a 
semi-classical one.  Hence we must define a class of $C^*$ dynamical
systems for which it makes sense to
speak of the semi-classical limit.  To this end, we introduce in \S1 the
class of {\it quantized Gelfand-Segal} systems.  For such systems, there will be
a well-defined ``energy" 
$$E:\N_A\cap \E(E^G_\A)\rightarrow \R^+$$
on the set of normal ergodic states;  roughly speaking, to each such
state $\rho$ will correspond an irreducible $\sigma \in \hat U$ and the 
energy will be defined by
$$E(\rho) = \delta (\sigma ,\upharpoonright) $$
with $\delta(\sigma ,\upharpoonright)$  more or less the distance of
$\sigma $ from the trivial representation of $G$.  Above $\hat U$ is the
spectrum of $U$, i.e.\ the set of irreducibles $\sigma $ in the unitary dual
$\hat G$ of $G$ which occur in $U$.  Moreover, there will exist for each
$E>0$ a well-defined {\it microcanonical ensemble} $\omega _E$ at
energy level $E$, which will essentially be the average of all normal
ergodic states $\rho$ of energy $E(\rho)\leq E$.  Enough (in fact, more
than enough) will be assumed about $\hat G$ and $\hat U$ to make the
definitions of $E$ and $\omega _E$ run smoothly.

The key property of quantized Gelfand-Segal systems will be the following:
\begin{itemize}
\item there exists a unique ``classical limit" state $\omega $ such that
$\omega _E\rightarrow \omega $ weakly as $E\rightarrow \infty$.
\end{itemize}
By the Gelfand-Segal construction (\S1; [B.R]), $\omega $ gives rise to a 
 a cyclic representation $\pi_{\omega}$ of $\A$, and a unitary representation $U_{\omega}$
of G, on a Hilbert space $\Hh _{\omega}$.  As mentioned above, the induced system
$(\pi_{\omega}(\A), G, \alpha_{\omega})$ will play the role of the classical limit, and 
$(\A,G,\alpha)$ will be regarded as its quantization.  Of course, the classical limit need
not be abelian;  if it is,  the original system will be called  {\it quantized abelian}.  
For the proofs of Theorems 1 and 2 it is
in fact sufficient that the pair $(\A, \omega)$ be ``G-abelian" (see [B.R] or \S1 for 
the definition).  In this case the original system will be called {\it quantized G-abelian}.

To illustrate the notion of quantized abelian system, consider the
 example above with $G=\R$, $\Hh =
L^2(M)$ and $U_t=\expit \sd$.  The relevant algebra is $\A =
\Psi^\circ(M)$, the algebra of zero-th order pseudodifferential operators
on $M$ (or its $C^*$ closure, to be perfectly precise).  The action of $\R$
is given by $\alpha _t(A) = U^*_tAU_t$.  The spectrum of $U$ is of
course the set of characters $\{\expit \sqrt{\lambda _j}\}$, and $\delta
(\expit \sqrt{\lambda _j}, \upharpoonright)=\sqrt{\lambda _j}$.  The
normal ergodic states are given by $\rho_j (A) = (A\varphi_j,
\varphi_j)$, and the energy $E(\rho_j) =\sqrt{\lambda _j}$.  The
microcanonical ensemble is 
$$\omega _E =\frac{1}{ N(E)}\sum_{\sqrt{\lambda _j} \leq E} \rho_j$$
and as is well-known it tends to the  state
$$\omega (A) = \int_{S^*M} \sa d\mu\;.$$  The
classical limit system is then $G^t$ acting on $ L^2(S^*M,d\mu)$; hence
the original system is quantized abelian.  For further
discussion, see \S3.

Postponing the precise definitions until \S1, we can state our main
abstract result as follows:

\begin{thone}  Let $(\A,G,\alpha )$ be a quantized abelian (or G-abelian) system and
suppose that the classical limit state $\omega $ is an ergodic
state.

Then, for any admissible density $D^*$ on the set $\N_A\cap \E (E^G_\A)$ of normal
ergodic states, there exists a subset ${\cal S} \subset \N_A\cap \E
(E^G_\A)$ such that:

(a)  $D^* ({\cal S}) = 1$

(b)  $\displaystyle{\weaklim\begin{Sb} E(\rho)\rightarrow
\infty\\ \rho\in {\cal S}\end{Sb}} \rho = \omega $. \end{thone}

For the previous example of $(\Psi^\circ (M), \R,\alpha )$, the theorem
shows that
$$(A\varphi_j,\varphi_j) \rightarrow
\int_{S^{*}M}\sa d\mu \quad\quad (\lambda _j\in {\cal S})$$
where ${\cal S}$ is a subset of full counting density in the spectrum
$\{\lambda _j\}$, and $A\in \Psi^\circ(M)$.  Hence Theorem 1 gives a
rather abstract version of the quantum ergodicity theorem that
eigenstates of quantizations of classical ergodic systems become
uniformly distributed on energy surfaces in the high energy limit
([Sn],[CV],[Z.1]).

The proof of Theorem 1 is quite simple, and indeed simplifies the
previous proofs.  The underlying idea (which is perhaps not visible in the
proof) is even simpler:  By assumption, the limit state $\omega $ is an
extreme point of the compact convex set of invariant states.  The
condition $\omega _E\rightarrow \omega $ states more or less that
$\omega $ is the barycenter of the set of pure normal invariant states. 
This is a contradiction unless these pure states tend individually to
$\omega $.  This idea suggests that Theorem 1 may admit a more general
formulation.  In the actual proof, the additional fact is used that for ergodic states
of abelian  systems,or  for ergodic G-abelian pairs, there is "uniqueness of the vacuum" 
in the associated
classical limit (i.e. rank $E_{\omega}$ = 1; see [R, p. 155] for terminology).

The conclusion of Theorem 1 may be taken as a definition of the quantum
ergodicity of a quantized abelian or G-abelian system.  To obtain a better
understanding of it, we  reformulate it in terms of the operator
averages
$$\langle A\rangle_\alpha  = \int_G\chi_\alpha (g) \alpha _g (A) dg$$
where $\{\chi_\alpha \}$ is an ``$M$-net" for the amenable group
$G [R]$.  For instance, if $G=\R^n\times T^m\times \Z^k\times K$ as
above, then $\chi_\alpha (g)dg$ could be the product
$\chi^{\R^{n}}_\alpha (x)dx\otimes d\theta \otimes
\chi^{\R^{k}}_\alpha (n)dn \otimes d\mu$, where $d\mu$ (resp.\
$d\theta$) is the normalized Haar measure on $K$ (resp.\ $T^n$),
where $\chi^{\R^{n}}_\alpha $ is $\alpha ^{-n}$ times the
characteristic function of a cube of side $\alpha $ and where $dx$
(resp.\ $dn$) is Lesbesque measure (resp.\ counting measure on
$\Z^k$).   The limit as $\alpha \rightarrow \infty$ of $\langle A\rangle
_\alpha $ does not exist in $\A$, but it does exist in the $W^*$
(von Neumann) closure of $\pi(\A)$, i.e.\ the closure in the strong
operator topology of ${\cal L}(\Hh)$.  We will denote this closure of
$\A$ by $\M$, and set
$$\langle A\rangle =\operatornamewithlimits{w
-\lim}_{\alpha \rightarrow \infty}\langle A\rangle_\alpha \;.$$
Following [Su] and [Z.3], we will say:

\subsection*{(0.1) Definition}  Let $(\A,G,\alpha )$ be a quantized GS
 system.  Say, $(\A,G,\alpha )$ is a {\it quantum
ergodic} system, if there exists an (invariant) state $\omega \in E^G_\A$
such that for all $A\in \A$,
$$\langle A\rangle =\omega (A) I+K$$
where $K\in \M$ and where
$$\lim_{E\rightarrow \infty}\omega _E(K^*K)=0\;.$$

Thus, the time average of an observable equals its space average
plus an asymptotically neglible error as $E\rightarrow \infty$. 
Note that $\omega _E$ is normal, so is well-defined on $K^*K$. 
  We have

\begin{thtwo} Let $(\A,G,\alpha )$ satisfy the assumptions of
Theorem 1.  Then it is a quantum ergodic system.\end{thtwo}

We remark that the state $\omega $ in the definition of quantum
ergodicity is necessarily the $\weak^*$ limit of $\omega _E$. 
However, it is not clear that it has to be ergodic; there may exist
quantum ergodic systems which are not classically ergodic.  Regarding
this converse direction we have the following result (cf. [Su][Z.2,5]):

\begin{ththree} Let $(\A, G, \alpha)$ be a quantized abelian system, with
G abelian. 

(a) Suppose $\omega$ is ergodic. Then 
$$ \lim_{T \rightarrow \infty}\lim_{E \rightarrow \infty} \omega_E (<A>_T^*A) =
\lim_{E \rightarrow \infty} \lim_{T \rightarrow \infty} \omega_E(<A>_T^*A)
=|\omega(A)|^2 \leqno(0.2).$$

(b) Suppose conversely that $(\A, G, \alpha)$ is quantum ergodic and (0.2) holds.  Then $\omega,$
hence the classical limit system, is ergodic. \end{ththree}

The condition (0.2) is of course equivalent to
$$\lim_{T\rightarrow \infty} \omega(<A>_T^* A)=\lim_{E \rightarrow \infty}\omega_E(<A>^*A)
=\omega(<A>^*A) \leqno(0.3)$$
at least when $\omega$ extends to the von-Neumann completion.  Less obviously it is 
equivalent to 
$$\forall \epsilon \exists \delta \lim_{E \rightarrow \infty}\frac{1}{N(E)}
 \sum_{i\not=j, |\chi_i|,|\chi_j| \leq E,|\chi_i - \chi_j| \leq \delta}
 |(A \phi_i,\phi_j)|^2 \leq \epsilon \leqno(0.4)$$
where $\{\phi_i\}$ is an orthonormal basis of joint eigenfunctions and where
$\{\chi_i\}$ are the corresponding eigenvalues (characters) (cf. [Su][Z.2]).
The proof of (b) is based on the following

\subsection*{Spectral measure Lemma } {\it Define the measure $dm_A$ on $C_c(\hat G)$ by
$$\int_{\hat G} f(\chi) dm_A(\chi):= \lim_{E \rightarrow \infty} \omega_E(<A>_{
{\cal F}f}^*A) \leqno(0.3)$$ with ${\cal F}f$ the Fourier transform of $f$ and
with $<A>_h: = \int_G h(g) \alpha_g(A) dg.$  Then: $dm_A$ is the spectral measure for the
classical dynamical system corresponding to vector $\pi_{\omega}(A)$.}

Theorems 1 and 2 have a number of applications to $C^*$ dynamical
systems $(\A,G,\alpha )$ where $\A$ is an algebra of
pseudodifferential, Fourier Integral or Toeplitz operators.  We will
present some rather simple examples with $G=\R$ in \S3;  more elaborate examples
will be presented in [Z.Zw] and [Z.5].

\subsection*{Acknowledgements}  We have profited from discussions with F.Klopp and
M.Zworski.  The billiards example is a by-product of [Z.Zw].

\section {Quantized Gelfand-Segal systems}

In this section we state more precisely the conditions on $(\A, G,\alpha
)$ which are assumed in the statements of Theorems 1 and 2.

As mentioned above, $\A$ will be assumed to be unital and separable and
$(\A, G, \alpha )$ will be assumed to have an effective covariant representation
$(\Hh,\pi,U)$ on a Hilbert space $\Hh$.   ${\cal A}$ will also be assumed to contain
a subalgebra ${\cal K}$ which gets represented as the compact operators on $\Hh.$
We will further assume that the spectrum $\hat U$ of $U$ is discrete, in particular that the
multiplicity $m(\sigma)$ of each $\Hh_\sigma $ is finite.     We then denote by
$$\Hh = \bigoplus_{\sigma \in \hat U} \Hh_\sigma $$
the isotypic decomposition of $U$ and by
$$\Pi_\sigma : \Hh\rightarrow \Hh_\sigma$$
the orthogonal projection onto the isotypic summand $H_\sigma $.  

For the sake of simplicity, we will assume that $G$ is an amenable Lie
group of the form
$$G = \R^n\times T^m \times \Z^k \times K\;,$$
where $T^m$ is the real $m$-torus and where $K$ is a compact
semi-simple Lie group.  Hence the unitary dual $\hat G$ of $G$ has the
form 
$$\hat G = \R^n \times \Z^m\times T^k\times \hat K$$
where in the usual way we identify
$$\hat K = I^* \cap {\frak t}^*_+$$
with $I^*$ the lattice of integral forms and ${\frak t}^*_+$ a closed Weyl
chamber in the dual of a Cartan subalgebra.  We can then define the
``distance" $\delta (\upharpoonright,\sigma )$ of a representation
$\sigma \in \hat G$ from the trivial representation by
$$\delta (\sigma,\upharpoonright ) = |\bar \sigma |$$
where $|\cdot|$ is the Euclidean norm on $\R^n\times\R^m \times {\frak t}^*_+$, and
where $\bar \sigma $ is the projection of $\sigma $ to this space.  We
regard $\delta(\sigma,\upharpoonright )$ as the semi-classical
parameter, i.e.\ as the inverse Planck constant or energy level.

The numerical spectrum
$$spec(U):= \{ \delta (\sigma,\upharpoonright ): \sigma \in \hat U \} $$
of ``energy levels" is then a discrete subset of $\R^+$,
$$0=E_o < E_1 < E_2 < \dots \uparrow \infty.$$
There are two natural notions of the multiplicity of an energy level.  The
first,  given  by
$$m(E_j) = \sum_{\sigma \in \hat U, \delta(\sigma, \upharpoonright)=E_j} rank\Pi_{\sigma},$$
counts the total dimension in the energy range, while the second
$$m^*(E_j)=\sum_{\sigma \in \hat U, \delta(\sigma, \upharpoonright)=E_j} m({\sigma})$$
counts the number of irreducibles.  They give rise to
the two spectral counting functions  
$$N(E):= \sum_{j: E_j \leq E} m(E_j),$$
respectively
$$N^*(E):= \sum_{j: E_j \leq E} m^*(E_j).$$
In many applications, $N(E)$ has an asymptotic expansion as $E \rightarrow \infty$, and
$m(E_j)$ is of strictly lower order than $N(E_j)$, but it does not seem natural in the rather
general context of this section to introduce too many hypothesis on the spectrum. 
To avoid pathologies, however, we will assume that the $\spec(U)$ is {\it regular} in the sense
that
$$m(E_{j+1}) \leq C N(E_j),\;\;\;\;\;\;\;\;\;\;\;\;\;\;\;\;\;\;m^*(E_{j+1})\leq C N^*(E_j)$$
for some $C > 0.$

Corresponding to each isotypic summand $\Hh_{\sigma}, \sigma \in \hat U$ we define the normal
invariant state
$$\omega _\sigma (A) = \frac{1}{\rank \Pi_\sigma }\Tr\Pi_\sigma 
A\;.$$
Note that $\omega _\sigma $  is not
ergodic unless the multiplicity of $\sigma $ in $\Hh_\sigma $ is one:  In
fact, the normal ergodic states are in one-one correspondence with
projections $P$ onto irreducible subspaces in $\Hh$.  To see this, recall that a normal
invariant state corresponds to a density matrix (positive trace-class
operator) $\rho$ which commutes with $G$.  It is therefore a sum of
scalar multiples of projections onto irreducibles, and is indecomposible
if and only if it is a multiple of one such projection.  Since it has unit
mass, each normal ergodic state $\rho$ must be of the form $\rho(A) =
\frac{1}{d(\sigma )} \Tr P_\sigma A$ where $d(\sigma)  = \Tr
P_\sigma $ and $P_\sigma $ is a projection onto an irreducible subspace
of some type $\sigma \in \hat U$.

We then introduce the {\it microcanonical ensemble at energy level E},
$$\omega _E:=\frac{1}{N(E)}\sum _{\sigma :\delta(\sigma
,\upharpoonright)\leq E}(\rank \Pi_\sigma ) \omega _\sigma. $$
It is the state corresponding to the usual microcanonical density matrix
$$\frac{1}{N(E)}\sum_{\sigma: \delta(\sigma, \upharpoonright)\leq E} \Pi_{\sigma}$$
(see [T,(2.3.1)], and is the most mixed combination of the states of energy less than E.
We also introduce the ensemble
$$\tilde{\omega}_E:= \frac{1}{N^*(E)}\sum_{\sigma: \delta(\sigma, \upharpoonright)\leq E} 
m(\sigma) \omega _\sigma$$
 with
$$N^*(E):=\sum_{\delta(\sigma ,\upharpoonright)\leq E} m(\sigma).$$
which is the most mixed combination of the normal  ergodic states of energy less than E.
 Both ensembles seem to be natural candidates for the microcanonical ensemble, and
 the statements of Theorems 1-3 are  valid for both.
Since $\omega_E$ differs from $\tilde{\omega}_E$ only in weighting the $\sigma$th
term by $d(\sigma)$, the two ensembles coincide if $G$ is abelian, and in 
   applications  the two ensemble averages are often asymptotically equivalent, i.e. 
$\omega_E(A) \sim \tilde{\omega_E}(A)$.  

Associated to the microcanonical ensemble $\omega_E$ (or $\tilde{\omega_E}$)
is the corresponding  collection 
 of {\it admissible densities} on the set of normal ergodic states.
To define these densities,
we denote by $S_\sigma $ the set of irreducible subspaces in
$\Hh_\sigma $, and by $\N_\A \cap \E(E^G_\A)$ the set of normal
ergodic states.  We then have 
$$\N_A \cap \E (E^G_\A) = \bigsqcup_{\sigma \in\hat U}S_\sigma \;,$$
and define the {\it energy} of a normal ergodic state by
$$E(\rho) = \delta (\sigma ,\upharpoonright)\quad\quad (\rho\in
S_\sigma )\;.$$
The admissible densities $D^*_\nu$ on $\N_A \cap \E(E^G_\A)$ are constructed from
  families $\nu = \{\nu_\sigma  :\sigma \in \hat U\}$ of
unit mass measures on the $S_\sigma $'s, with each
$\nu_\sigma $ giving a barycentric decomposition 
$$\omega
_\sigma  = \int_{S_{\sigma }} \omega _\phi d\nu_\sigma (\phi)$$
 of $\omega _\sigma $ into ergodic states.  To define the corresponding
density, we note that a subset
${\cal S}\subset \N_A \cap \E (E^G_\A)$ has the form
$${\cal S} = \bigsqcup _{\sigma \in \hat U}\tilde{{\cal S}}_\sigma
\quad\quad(\tilde{{\cal S}}_\sigma  = {\cal S} \cap S_\sigma ).$$
For the choice $\omega_E$ of microcanonical ensemble, we then set
$$D^*_\nu ({\cal S}):=\lim_{E\rightarrow \infty} \frac{1}{N(E)}
\sum_{\delta(\sigma ,\upharpoonright)\leq E}\nu_\sigma (\tilde{{\cal S}}_\sigma
)\rank\Pi_\sigma .$$
In the case of $\tilde{\omega_E}$, we define the density $\tilde{D}^*_{\nu}$ analogously
but with $m(\sigma)$ in place of rank$\Pi_{\sigma}$.
 In the simplest case where $G$ is abelian and $U$ is multiplicity free, both
densities coincide and are given by
$D^*({\cal S}) = \lim_{E\rightarrow \infty}\frac{1}{N(E)}\# \{\sigma \in
{\cal S}:\delta (\sigma ,\upharpoonright) \leq E\}$.  

We then say:

\subsection*{(1.1) Definition} $(\A,G,\alpha )$ is a {\it quantized Gelfand-Segal
system} if it satisfies the following conditions:

(a)  $G = \R^n\times T^m\times \Z^k\times K$ ;

(b)  $\hat U$ is discrete and $spec(U)$ is regular;

(c)   There exists an invariant state $\omega$ such that
 $\lim_{E\rightarrow \infty} \omega _E = \omega $. 
\bigskip

In (c), the limit is understood to be in the $\weak^*$ sense.  Corresponding to the
choice of $\tilde{\omega}_E$, (c) is of course replaced by

(c') There exists an invariant state such that $\lim_{E\rightarrow \infty}
\tilde{ \omega} _E = \omega $.

Let us recall that by the Gelfand-Segal (or GS)
construction [R; A.3.5, 6.2.2], [B.R], the invariant state
$\omega $ gives rise to a covariant cyclic representation $(\Hh_\omega ,
\pi_\omega ,U_\omega ,\Omega_\omega )$ of $(\A,G,\alpha )$ with the
properties  
\begin{gather*} \alpha _\omega (g)\pi_\omega (A) :=
U_\omega (g) \pi_\omega (A) U_\omega (g)^{-1} = \pi_\omega (\alpha
_g(A))\\ U_\omega (g) \Omega_\omega= \Omega_\omega \\
\omega (A) = (\Omega_\omega , \pi_\omega (A) \Omega_\omega
)\;.\end{gather*}
We recall that the Hilbert space $\Hh_\omega $ is the closure of
$\A/\N$ with respect to the inner product $\omega (AB^*)$, where $\N$
is the left ideal $\{A\in \A:\omega (A^*A) = 0\}$.  Also, that the
representation $\pi_\omega $ is defined by $\pi_\omega (A)(B+\N) =
(AB+\N)$; that $\Omega_\omega  = I+\N$; and that $U_\omega (g)(B+\N) = (\alpha _g
(B)+\N)$.  The new $C^*$ dynamical system $(\pi_{\omega}(\A), G, \alpha_{\omega})$
will be referred to as the classical limit of $(\A, G, \alpha).$

 In  semi-classical analysis, it is natural to focus on
the case where  $\pi_\omega (\A)$ is abelian,
and hence isomorphic to $C(X)$ for a compact Hausdorff space $X$.  We
recall that $X$ is the set of pure states of $\pi_\omega (\A)$, and
that the isomorphism is given by $A+\N\rightarrow \psi_A$,
where $\psi_A(\rho) = \rho (A+\N)$.  As the notation suggests
$\psi_A$ will denote the element of $C(X)$ corresponding to $A$
under the composition $A\rightarrow \pi_\omega (A)\rightarrow
\psi_A$.  Also, it is clear that the states of $\pi_\omega (A)$
determine states of $\A$ which annihilate $\N$.   Under this isomorphism,
the states of $\pi_\omega (\A)$ correspond to the probability measures
on $X$.  In particular, $\omega $ induces the state $\pi_\omega (A)
\rightarrow (\Omega_\omega , \pi_\omega (A)\Omega_\omega )$.  Let us
denote by $\mu$ the corresponding measure.  Then $\Hh_\omega
\simeq L^2(X,\mu)$, and the automorphisms $\alpha _\omega (g)$
determine a group of measure preserving transformations of
$(X,\mu)$ and the unitary group $U_\omega (g)$ of translations in
$L^2(X, \mu)$.  We will say: 

\subsection*{(1.2) Definition)} $(\A, G, \alpha)$ is a {\it quantized abelian} system if
it is a quantized GS system and if the classical limit system is abelian. 

It is potentially interesting to consider quantized GS systems with nonabelian  classical
limits.  For the purposes of this paper, a second natural condition on the classical limit
 is the uniqueness of the vacuum state.    We recall that this means that the projection
 $E_{\omega}$
onto the $U_{\omega}(G)-$ invariant vectors in $\Hh_{\omega}$ has rank one, i.e. that
$\Omega_{\omega}$ is the unique invariant vector up to scalar multiples.  This is equivalent
to ergodicity of $\omega$ (or equivalently of $\mu$) in the abelian case, or more generally 
in the case where the algebra generated by $E_{\omega}\pi_{\omega}(\A)E_{\omega}$
 is abelian (i.e. if $(\A, \omega)$ is a ``G-abelian pair", see [BR, Proposition 4.3.7 and
Theorem 4.3.17]).  Hence we also distinguish the following case:

\subsection*{(1.3) Definition} $(\A, G, \alpha)$ is a {\it quantized G-abelian} system if
it is a quantized GS system and if $(\A, \omega)$ is a G-abelian pair.

We note that the usual terminology ``G-abelian" applies to systems for which all invariant states
define G-abelian pairs; while  here quantized G-abelian refers only to the classical limit state
$\omega.$

\section{Quantum ergodicity theorems}

The purpose of this section is to prove Theorems 1-3.  The following lemma provides
a simple model for the somewhat more complicated situation of Theorem 1:

\subsection*{(2.1) Lemma} {\it Let $(A,G,\alpha )$ be a $C^*$ dynamical
system with $G$ an amenable group.  Let $\{\rho_j:j=1,2,3,\ldots\}$ be
any sequence of $G$-invariant states on $\A$, and let
$\rho_N=\frac{1}{N} \sum^N_{j=1} \rho_j$.
}\medskip

{\it Assume:

(a) 
$\displaystyle{\operatornamewithlimits{\weak^*-\lim}_{N\rightarrow
\infty}\rho_N}$ exists.

(b)  The Gelfand-Segal system defined by the limit $\omega $ has a unique vacuum state.}

\noindent {\it Then, there exists a subsequence ${\cal S}\subset\Nb$ of
counting density one such that}
$${\it \operatornamewithlimits{\weak^*\lim}\begin{Sb}
j\rightarrow \infty\\
j\in {\cal S}\end{Sb} \rho_j =\omega} \;.$$

\begin{pf}  Let $A\in \A$, and consider the sums
$$S_2 (N,A) = \frac{1}{N} \sum^N_{j=1}|\rho_j(A) - \omega (A)|^2\;.$$
Since $\rho_j$ is $G$-invariant,
$$S_2(N,A) = \frac{1}{N} \sum^N_{j=1} |\rho_j (\langle A\rangle
_\alpha ) - \omega (A)|^2\;.\leqno(2.2)$$
By the Schwartz inequality for positive linear functionals
([B.R.,Lemma 2.3.10]),
$$|\rho_j(\langle A\rangle_\alpha ) -\omega (A)|^2 =|\rho_j(\langle
A\rangle_\alpha -\omega (A))|^2 \leq \rho_j((\langle
A\rangle_\alpha -\omega (A))^*(\langle A\rangle_\alpha-\omega
(A)))\;.$$
Hence,
$$S_2(N,A)\leq \rho_N[(\langle A\rangle_\alpha-\omega
(A))^*(\langle A\rangle_\alpha-\omega (A))]\;.\leqno(2.3)$$
Letting $N\rightarrow \infty$ we obtain
$$\overline{\lim_{N\rightarrow \infty}} S_2 (N,A) \leq
\omega [(\langle A\rangle_\alpha-\omega (A))^*(\langle
A\rangle_\alpha-\omega (A)))]\;.\leqno(2.4)$$
We now claim:
$$\lim_{\alpha \rightarrow \infty} \omega [(\langle
A\rangle_\alpha-\omega (A))^*(\langle A\rangle_\alpha-\omega
(A))]=0\;.\leqno(2.5)$$
Indeed, (2.5) is equivalent to the condition rank $E_{\omega} = 1$ if G is amenable,
 see [R, Proposition 6.3.5].  
Hence, we have proved that for any $A\in \A$, 
$$\lim_{N\rightarrow \infty}\frac{1}{N} \sum^N_{j=1}
|\rho_j(A)-\omega (A)|^2 = 0\;.\leqno(2.6)$$
By a standard lemma on averages of positive numbers
[W,Theorem 1.20], (2.6) implies that for each $A\in \A$, there
is a subsequence ${\cal S}_A\subset \Nb$ of counting density one such that
$$\lim\begin{Sb} k\rightarrow \infty\\k\in {\cal S}_A\end{Sb}
\rho_k (A) = \omega (A)\;.\leqno(2.7)$$
To obtain a density one subsequence ${\cal S}$ independent of $A$,
we use a diagonalization argument ([CV], [Z.1]).  Since
$\A$ is separable, there exists a countable dense subset $\{A_j\}$
of the unit ball of $\A$.  For each $j$, let ${\cal S}_j\subset \Nb$ be a
density one subsequence such that (2.7) is correct for $A_j$.  We
may assume ${\cal S} _j\subset {\cal S}_{j+1}$.  Then choose $N_j$ so that 
$$\frac{1}{N}\#\{k \in {\cal S}_j:k\leq N\}\geq 1-2^{-j} \mbox{ for }
N\geq N_j\;.$$
Let ${\cal S}_\infty$ be the subsequence defined by
$${\cal S}_\infty \cap [N_j, N_{j+1}]={\cal S}_j \cap [N_j, N_{j+1}] \leqno(2.8).$$
Then ${\cal S}_\infty$ is of density one and
$$\lim\begin{Sb}k\rightarrow \infty\\k\in {\cal S}_{\infty}\end{Sb}
\rho_k (A) = \omega (A)\leqno(2.9)$$
for all $A\in \A$: as follows since (2.9) holds for the set
$\{A_j\}$ and since $\{A_j\}$ is dense in the unit ball. \end{pf}

\subsection*{(2.10) Remark 1} Uniqueness of the vacuum state implies that  $\omega$
is an ergodic state [R, Theorem 6.3.3].  It is equivalent to ergodicity of $\omega$ if
the pair $(\A, \omega)$ is G-abelian [loc.cit].  In particular, if the GS system is
abelian, it is equivalent to ergodicity of the induced flow.  Hence: 

\subsection*{(2.11a) Corollary} {\it The conclusion of Lemma (1.2) is
correct if we replace assumption (b) with the assumption that $(\A, \omega)$ is abelian
and that $\omega$ is ergodic.}\medskip

\subsection*{(2.11b) Corollary} {\it The conclusion of Lemma (1.2) if in
place of (b) we assume $\omega $ is ergodic and $(\A,\omega )$ is
$G$-abelian. }\medskip

We now give the \medskip

\subsection*{Proof of Theorem 1} Let us consider first the special case
$$\nu_\sigma   = \frac{1}{m(\sigma )}\sum^{m(\sigma )}_{j=1}
 \delta_{\omega\sigma j}\;.\leqno(2.12)$$
Here $\delta_{\omega_{ \sigma j}}$ is the point mass at the ergodic state
$$\omega_{ \sigma j}(A) = \frac{1}{d(\sigma )} \Tr\Pi_{\sigma j} A
$$
where we have chosen a decomposition
$$\Pi_\sigma  = \bigoplus ^{m(\sigma )}_{j=1} \Pi_{\sigma j}$$
corresonding to a decomposition $\Hh_\sigma =\bigoplus ^{n(\sigma
)}_{j=1} \Hh_{\sigma j} $ of $\Hh_\sigma $ into irreducibles.  Also, we
recall that $d(\sigma )$ is the dimension of the irreducible and $m(\sigma )$ is its
multiplicity in $\hat U$.  The
 associated density
$D_{\nu}^*$ is then supported on the set $\{\omega_{\sigma j}\}$.

Conside first the choice of $\tilde{\omega}_E$ as microcanonical ensemble, since it
is somewhat simpler to work with.  We have, by (1.1c'),
 $$\weaklim_{E\rightarrow \infty}\frac{1}{N^*(E)}
\sum\begin{Sb}\sigma: \\
\delta(\sigma ,1)\leq E\end{Sb}  \sum^{m(\sigma )}_{j=1}
\omega _{\sigma j} = \omega\leqno(2.13a) $$
for some  ergodic state $\omega $.  By the argument of Lemma 2.1 (leading to (2.6)),
we then have for each $A \in {\cal A}$
$$\lim_{E\rightarrow \infty}\frac{1}{N^*(E)} \sum\begin{Sb}\sigma:\\
\delta(\sigma ,1)\leq E\end{Sb}  \sum^{m(\sigma )}_{j=1}
|\omega _{\sigma j}(A) -\omega(A)|^2 = 0.\leqno(2.13b) $$ Our aim is then to construct a subset
$${\cal S} \subset \{\omega_{\sigma j}\}$$
with $\tilde{D}_{\nu}^*({\cal S})=1$ and such that
$$w-\lim\begin{Sb}E(\omega_{\sigma j}) \rightarrow \infty \\ \omega_{\sigma j} \in
{\cal S} \end{Sb}\omega_{\sigma j} = \omega \;.$$  We will in fact construct such a subset
of full counting density in a natural sense.

 We begin by arranging the ergodic states $\omega _{\sigma j}$ in a
sequence: First fix an ordering $\{\sigma _\ell:
\ell=1,2,\ldots\}$ of the irreducibles $\sigma $ occurring in $\hat
U$, with $\delta(\sigma_l,\upharpoonright) \leq \delta(\sigma_m, \upharpoonright)$
if $ l\leq m$,
 and then arrange the states $\omega _{\sigma _{\ell}j}$ in
lexicographic order, $\omega _{\sigma _{\ell}j}=\omega
_{n(\ell,j,)}$.  Henceforth we denote this sequence of states by $\{ \omega_n\}$.
  We also define postive integers
$N^*_m$ by $N^*_m:= N^*(E_m)$, where $\{E_m\}=spec(U)$.  We then have:
$$\lim_{N^*_m \rightarrow \infty}
\frac{1}{N^*_m}\sum_{n=1}^{N^*_m} |\omega_n(A) - \omega(A)|^2=0.$$  We note that
$$\frac{N^*_{m+1}}{N^*_m} = \frac{N^*_m + m^*(E_{m+1})}{N^*_m} \leq (1+C) $$
by regularity of the spectrum.  It follows that for $N^*_m \leq N \leq N^*_{m+1}$
$$\frac{1}{N} \sum_{n=1}^N |\omega_n(A) - \omega(A)|^2 \leq 
\frac{1}{N^*_m}\sum_{n=1}^{N^*_{m+1}} |\omega_n(A) - \omega(A)|^2$$
$$\leq (1+C)\frac{1}{N^*_{m+1}}\sum_{n=1}^{N^*_{m+1}} |\omega_n(A) - \omega(A)|^2$$
hence
$$\lim_{N\rightarrow \infty}\frac{1}{N} \sum_{n=1}^N |\omega_n(A) - \omega(A)|^2 = 0.
\leqno(2.13c)$$
As in the proof of Lemma 2.1, this implies the existence of a subsequence ${\cal S}_1
\subset \{\omega_n\}$ of counting density one such that 
$$w-\lim\begin{Sb}n \rightarrow \infty \\ \omega_n \in
{\cal S}_1 \end{Sb}\omega_n = \omega \;.$$

The choice of $\omega_E$ as microcanonical ensemble leads to the
  somewhat more complicated limit formulae
$$\lim_{E\rightarrow \infty}\frac{1}{N(E)} \sum\begin{Sb}\sigma:\\
\delta(\sigma ,1)\leq E\end{Sb} d(\sigma) \sum^{m(\sigma )}_{j=1}
|\omega _{\sigma j}(A) -\omega(A)|^2 = 0.\leqno(2.14a)$$ 
Ordering the states as above, and letting $d(n)$ denote the dimension of the representation
corresponding to $\omega_n$ we now have
$$\lim_{N^*_m \rightarrow \infty}
\frac{1}{N_m}\sum_{n=1}^{N^*_m}d(n) |\omega_n(A) - \omega(A)|^2=0.\leqno(2.14b)$$ 
where  $N_m:= N(E_m)=\sum_{n=1}^{N^*}d(n)$.  The regularity of the spectrum then
implies
$$\lim_{N\rightarrow \infty} \frac{1}{D(N)}\sum_{n=1}^{N} d(n)|\omega_n(A) -\omega(A)|^2
=0, \leqno(2.14c)$$ with $D(N):= \sum_{n=1}^N d(n).$  This leads to the conclusion that
there exists a subsequence ${\cal S}_1$ of $D^*$-density one of the set $\{\omega_n\}$
which tends to $\omega$ in the sense that
$$\lim_{N\rightarrow \infty} \frac{1}{D(N)} \sum \begin{Sb} n \leq N \\ \omega_n \in
{\cal S}_1\end{Sb} d(n) = 0.$$  
 With further hypotheses  on the distribution of irreducibles of $G$ in $\hat U$
and on the growth rate of the spectrum, this conclusion could be sharpened to give a
subsequence of counting density one tending to $\omega$ as in the case of $\tilde{\omega}_E$.
However, such hypotheses seem best left to arise naturally in applications.

  We now turn to
  the case of a general admissible density $D_{\nu}^*$ for $\omega_E$, for which we will
prove the existence of a subset of normal ergodic states of $D_{\nu}^*$-density one tending
to $\omega.$ In this general case, we have
$$\omega _E=\frac{1}{N(E)} \sum\begin{Sb}\sigma
:\\\delta(\sigma ,1) \leq E\end{Sb} \rank  \Pi_\sigma 
\int_{S_{\sigma }}\omega_ \phi d\nu_\sigma (\phi)\;.$$
Imitating the proof of Lemma 2.1 we let
$$S_2(E,A) = \frac{1}{N(E)}\sum\begin{Sb} \sigma :\\\delta
(\sigma ,1) \leq E\end{Sb}\rank \Pi_\sigma \int_{S_{\sigma }}|\omega_
\phi (A) - \omega (A)|^2 d\nu_\sigma (\phi)\;,\leqno(2.15a)$$
and by a similar argument obtain,
$$\lim_{E\rightarrow \infty}S_2 (E,A) = 0\;.\leqno(2.15b)$$
Our goal is  to construct a subset ${\cal S}\subset \N_A\cap \E (E^G_\A)$ of
$D_{\nu}^*$-density one such that
$\displaystyle{\lim\begin{Sb}  E(\phi)\rightarrow \infty\\\phi\in
{\cal S}_\infty\end{Sb}} \omega _\phi (A) = \omega (A)$ (all $A\in
\A$).

As in the proof of Lemma 2.1, we
 let $\{A_j\}$ denote a countable dense subset of the unit ball of $\A$ and begin
by constructing 
for each $A_j$ a subset ${\cal S}_j\subset \N_A\cap \E
(E^G_\A)$ such that $D^*_{\nu} ({\cal S}_j) = 1$ and such that
$$\lim
\begin{Sb} 
\phi\in{\cal S}_j\\E(\phi)\rightarrow \infty\end{Sb} \omega _\phi
(A_j) = \omega (A_j)\;.\leqno(2.16)$$
The construction is similar to that in the lemma on sequences.  We
let
$$J_{\sigma jk}=\{\omega_{\phi}\in S_\sigma :|\omega_ \phi (A_j) - \omega
(A_j)|^2>\frac{1}{k}\}\;.$$
and let $J_{jk}=\bigsqcup _{\sigma \in\hat U} J_{\sigma
jk}$.  Since
$$\int_{S_{\sigma }} |\omega_ \phi (A_j) -\omega (A_j)|^2
d\nu_\sigma (\phi) \geq \frac{1}{k} \nu_\sigma  \{|\omega_ \phi
(A_j) - \omega (A_j)|^2>\frac{1}{k}\}\leqno(2.17)$$
it is clear that $D^*_\nu(J_{jk}) = 0$ for all $jk$.  Hence there exist
integers $0 = \ell_0<\ell_1\leq \ell_2<\cdots$ such that for
$E\geq \ell_k$,
$$\frac{1}{N(E)} \sum\begin{Sb}  \sigma \in \hat U\\\delta
(\sigma ,1) \leq E\end{Sb} \rank\Pi_\sigma  \nu_\sigma
(J_{\sigma j(k+1)})<\frac{1}{k+1}\;.\leqno(2.18)$$
Let $J_j = \bigcup^\infty_{k=0} \displaystyle{\bigcup\begin{Sb}  \sigma
\in \hat U\\\ell_k <\delta (\sigma ,1) \leq \ell_{k+1}\end{Sb}}
J_{\sigma j(k+1)}$.  We claim that $D^*_\nu (J_j) = 0$.  Indeed,
$J_{\sigma jk}$ increases with $k$, so if $\ell_k \leq E\leq
\ell_{k+1}$, then 
$$\frac{1}{N(E)} \sum\begin{Sb} 
\sigma \in \hat U\\\delta (\sigma ,1) \leq E\end{Sb}
\rank\Pi_\sigma \nu_\sigma (J_j)\leq \frac{1}{N(E)} \sum\begin{Sb} 
\sigma \in \hat U\\\delta (\sigma ,1) \leq \ell_k\end{Sb}
\rank\Pi_\sigma \nu_\sigma (J_{\sigma j(k)})\leqno(2.19)$$
$$ +\frac {1}{N(E)} \sum\begin{Sb}  \sigma \in
\hat U\\\ell_k\leq\delta (\sigma ,1) \leq E\end{Sb}\rank\Pi_\sigma
\nu_\sigma (J_{\sigma j(k+1)})
 \leq \frac{1}{k} +
\frac{1}{k+1}\;.$$
Hence $\displaystyle{\lim_{E\rightarrow \infty}\frac{1}{N(E)}
\sum\begin{Sb}  \sigma \in \hat U\\\delta (\sigma ,1) \leq
E\end{Sb}}\rank\Pi_\sigma \nu_\sigma (J_j) = 0$.  Let ${\cal S}_j$ be
the complement of $J_j$.  We observe that 
$$\lim\begin{Sb} 
E(\phi)\rightarrow \infty\\
\omega_{\phi} \in {\cal S}_j\end{Sb}\omega _\phi
(A_j) = \omega (A)\;.\leqno(2.20)$$ 
Indeed, if $E(\phi)>\ell_k$ and
$\omega_{\phi}\not\in J_j$, then  $\omega_{\phi}\not\in J_{\sigma j(k+1)}$, and so
$|\omega _\phi(A_j) - \omega (A_j) |^2<\frac{1}{k}$.

Finally, we use a diagonal argument as in the proof of Lemma 2.1
to get rid of the dependence of ${\cal S}_j$ on $A_j$.  We may again
assume ${\cal S}_j\subset{\cal S}_{j+1}$, and choose $N_j$ so that
$$\frac{1}{N(E)}\sum\begin{Sb}  \sigma \in\hat U\\
\delta(\sigma ,1) \leq E\end{Sb}\rank\Pi_\sigma  \nu_\sigma
({\cal S}_{\sigma j} )\geq 1-2^{-j} \quad (E\geq N_j)$$
where ${\cal S}_{\sigma j} = {\cal S}_j\cap S_\sigma $.  We define
${\cal S}:={\cal S}_\infty$ by:
$${\cal S}_\infty\cap \bigcup\begin{Sb}  \sigma \in\hat U\\
N_j\leq \delta (\sigma ,1) \leq N_{j+1}\end{Sb} S_\sigma  =
\bigcup \begin{Sb}  \sigma \in \hat U\\N_j \leq \delta (\sigma
,1) \leq N_{j+1}\end{Sb}{\cal S}_{\sigma j}\;.$$
Then $D^*_\nu ({\cal S}_\infty) = 1$ and by a density argument
$\displaystyle{\lim\begin{Sb}  E(\phi)\rightarrow \infty\\\omega_{\phi}\in
{\cal S}_\infty\end{Sb}} \omega _\phi (A) = \omega (A)$ (all $A\in
\A$).   \qed 

\subsection*{(2.21) Remark} In the preceding, we let $\omega _E$ be the
average over the whole ``ball" of normal ergodic states of energy $\leq E$. 
But analogous results hold if we only average along a ray of
representations.  Such rays are frequently used to define
semi-classical limits.  So we include:

\subsection*{(2.22)  Addendum to Theorem 1 (Localized version)} {\it Let
$L$ be a ray of representations in $\hat G$, and let 
$$\Hh_L=\bigoplus_{\sigma \in \hat U\cap L} \Hh_\sigma \;.$$
Also let
$$\omega ^L_E: = \frac{1}{N(E,L)}\sum\begin{Sb}  \rho(\sigma
,1) \leq E\\
\sigma \in \hat U \cap L\end{Sb} \rank \Pi_\sigma \omega
_\sigma $$
with $N(E,L) = \sum\begin{Sb}  \sigma \in \hat U\cap L\\
\rho (\sigma ,1)\leq E\end{Sb}\rank \Pi_\sigma $}.\medskip

{\it Suppose $\displaystyle{\weaklim_{E\rightarrow \infty} }\omega ^L_E$
exists; let us denote it by $\omega ^L$.  Let $\N ^L_A\cap
\E(E^G_A)$ denote the set of normal ergodic states which occur in
the covariant representation $(\Hh_L,G,U|_{\Hh_{L}})$.}

{\it Then we have:  if  $(\A, \omega^L)$ is G-abelian, and $\omega ^L$ is ergodic, 
there is a subset ${\cal S}_L
\subset\N^L_A \cap \E(E^G_A)$ of relative density one such that
$$\weaklim\begin{Sb}  \rho\in {\cal S}_L\\E(\rho) \rightarrow
\infty\end{Sb}\rho = \omega ^L\;.$$}
Here, relative density one is as above with a set $\nu = \{
\nu_\sigma :\sigma \in\hat U \cap L\}$ of barycentric
decompositions of $\omega _\sigma $ for $\sigma \in L$.

The proof is essentially the same as for the full set $\hat U$, so
we omit it.  

We now give

\subsection*{Proof of Theorem 2.}  
We must show:
$$\lim_{E\rightarrow \infty}\omega _E[(\langle A\rangle-\omega
(A))^*(\langle A\rangle - \omega (A))]=0\;.\leqno(2.23)$$
We will see that this follows from a special case of Theorem 1. 
First, we observe that it is sufficient to prove it for $A$
satisfying $A^*=A$ or $A^* = -A$.  Indeed, we may express $A
=B+C$ with $B^*=B$, $C^*=-C$, and eliminate the cross term with
the Schwartz inequality $\omega (B^*C)^2\leq \omega (B^*B)\omega
(C^*C)$ for positive linear functionals.

Let us assume $A^*=A$ since the other case is similar.  Let us also first
assume that $G$ is abelian.  Then it is easily seen that
 $$\langle
A\rangle=\sum_{\sigma \in \hat U}\Pi_\sigma A\Pi_\sigma
\leqno(2.24)$$ so that
$$\begin{array}{lll}
\omega _\sigma  [(\langle A\rangle-\omega
(A))^2]&=&\frac{1}{\rank\Pi_\sigma }\Tr\Pi_\sigma (A-\omega
(A))\Pi_\sigma (A-\omega (A)),\\[12pt]
&=&\frac{1}{\rank\Pi_\sigma } \|\Pi_\sigma (A-\omega
(A))\Pi_\sigma \|^2_{\HS}\;,\end{array}$$
where $\|\cdot\|_{\HS}$ is the Hilbert Schmidt norm.  Since
$\Pi_\sigma (A-\omega (A))\Pi_\sigma $ is self-adjoint on
$\Hh_\sigma $, there exists an orthonormal basis $\{\phi_{\sigma
\ell}:\ell=1,\cdots, \dim\Hh_\sigma \}$ of its eigenvectors:
$$\Pi_\sigma (A-\omega (A))\Pi_\sigma \phi_{\sigma \ell} =
\langle (A-\omega (A))\phi_{\sigma \ell},\phi_{\sigma
\ell}\rangle\phi _{\sigma \ell}\;.$$
Hence,
$$\omega _\sigma  [(\langle A\rangle - \omega
(A))^2]=\frac{1}{\rank \Pi_\sigma }\sum^{\rank \Pi_{\sigma
}}_{\ell=1} |\langle (A-\omega  (A))\phi_{\sigma
\ell},\phi_{\sigma \ell}\rangle|^2\;.\leqno(2.25)$$
Let $\nu_\sigma  = \frac{1}{\rank\Pi_{\sigma }} \sum^{\rank
\Pi_{\sigma }}_{\ell=1} \delta _{\omega _{\sigma \ell}}$ where
$\omega _{\sigma \ell} (B) = \langle B\phi _{\sigma \ell},
\phi_{\sigma \ell}\rangle$.  Note that $\omega _{\sigma \ell}$ is
ergodic invariant state since $G$ is assumed to be abelian.  Hence
$$\omega _E [(\langle A\rangle-\omega (A))^* (\langle A\rangle -
\omega (A))] = S_2 (E,A)\quad\quad\text{(cf.\ (2.14))}\;,\leqno(2.26)$$
The conclusion now follows from (2.14a). 

Now let us consider $G = G_a\times K$ where $G_a$ is abelian and $K$ is
a compact Lie group.  We then have
$$\langle A\rangle=\sum_{\sigma \in \hat U} \Pi_\sigma \langle
A\rangle \Pi_\sigma$$
where $\Pi_\sigma \langle A\rangle\Pi_\sigma $ is an intertwining
operator from $\Hh_\sigma $ to itself.  Hence each eigenspace of
$\Pi_\sigma \langle A\rangle \Pi_\sigma $ is an invariant subspace, and
we have a spectral decomposition
$$\Pi_\sigma \langle A\rangle \Pi_\sigma  = \sum^{m(\sigma )}_{i=1}
\lambda _{\sigma i} \Pi_{\sigma i}$$
where $\Pi_{\sigma i}$ projects to an irreducible subspace.  The
eigenvalue is obviously given by, $\lambda _{\sigma i} = \omega _{\sigma i} (A)$. 
Hence,
$$\omega _\sigma [(\langle A\rangle - \omega (A))^2] =
\frac{1}{m(\sigma )}\sum^{m(\sigma )}_{i=1} |\omega _{\sigma i} (A) -
\omega (A)|^2\;.$$
Now let
$$\nu_\sigma  = \frac{1}{m(\sigma )}\sum^{m(\sigma )}_{i=1} \delta
\omega _{\sigma i}\;.$$
and apply (2.14-2.14a) as above.  \qed

Finally, we give

\subsection*{Proof of Theorem 3}  
Some general remarks before the proofs of (a) and (b) proper:
 Since the system is abelian, $U_{\omega}(g)$ is translation by an action of G by
measure-preserving transformations on $L^2(X, \mu)$ (\S1).  By definition, the spectral
measure for this action corresponding to  the vector $\psi_A \in L^2(X,\mu)$ is the measure
$d\mu_A$  on $\hat G$ defined by
$$(U_{\omega}(g) \psi_A, \psi_A) = \int_{\hat G} \chi(g) d\mu_A(\chi)  \leqno (2.27).$$
Here we identify $\hat G$ with the dual group of characters $\chi$ of $G$.  Ergodicity of
the action is then equivalent to the condition 
$$(d\mu_A - |\omega(A)|^2 \delta_1)(\{1\})=0 \leqno (2.28)$$ 
i.e. this measure has no point mass at the trivial character $1$ for any $A.$ 
 We may rewrite this condition in
terms of the invariant mean on $ G$ as follows:
$$\lim_{T \rightarrow \infty} \int_{ G} \int_{\hat G} \chi(g) (d\mu_A - |\omega(A)|^2
\delta_1)(\chi) \chi_T(g)dg = 0 \leqno (2.29).$$
Here as above $\chi_T$ denotes an M-net for the invariant mean on G, while $\chi$ alone
denotes a character of G. 

Temporarily assuming the Spectral measure Lemma, we now give the proofs of (a)-(b):

\noindent(a)  Since $\omega$ is ergodic, (2.29) holds.  From the Spectral measure Lemma,
we get
$$\lim_{T \rightarrow \infty} \int_G \int_{\hat G} \chi(g)(dm_A - |\omega(A)|^2\delta_1)(\chi) 
\chi_T(g)dg = 0 \leqno (2.30).$$
By the definition of $dm_A$ this gives
$$\lim_{T \rightarrow \infty} \lim_{E \rightarrow \infty}\omega_E(<A>_T^*A) = |\omega(A)|^2.$$
However, by Theorem 1, the right side is the same as the left side with the order of the limits
reversed.  Indeed, we have $<A> = \omega(A) I + K$ so
$$\lim_{E \rightarrow \infty} \lim_{T \rightarrow \infty} \omega_E (<A>_T^*A) =
\lim_{E \rightarrow \infty} \omega_E(<A>^*A)$$
$$=\lim_{E \rightarrow \infty}\omega_E((\omega(A) I + K)^*(A))= |\omega(A)|^2$$
by the Schwartz inequality. \qed

\noindent(b) Conversely, if the system is quantum ergodic and if (2.30) holds, then by
reversing the steps we conclude that (2.29) holds. Hence the classical system is ergodic. \qed

Last we give the

\subsection*{Proof of the Spectral measure Lemma}

By definition of quantized abelian we have
$$\lim_{E \rightarrow \infty}\omega_E(\alpha_g(A)^*A)=(U_{\omega}(g)\psi_A,\psi_A)$$
Suppose now that ${\cal F}f \in L^1(G).$ Since $\omega_E(\alpha_g(A)^*A) \in C_b(G)$ we have
$$\int_G{\cal F} f(g)\ lim_{E \rightarrow \infty} \omega_E(\alpha_g(A)^*A)dg = \lim_{E \rightarrow
\infty} \int_G {\cal F}f(g)\omega_E(\alpha_g(A)^*A)= \lim_{E \rightarrow \infty} \omega_E(<A>_f^*A)$$
$$=<U_{\omega}({\cal F}f) \psi_A, \psi_A>= \int_{\hat G}  f( d\mu_{\psi_A})$$
where $U_{\omega}(h) = \int_G h(g) U_{\omega}(g)dg.$ \qed

\section{Examples:  Continuous time systems $(G = \R)$}

In this section we will present four applications of Theorem 1 to
quantum ergodic systems $(\A, G,\alpha )$ with $G = \R$.  The algebras
$\A$ will be ($C^*$ closures of) $*$ algebras of Fourier Integral
operators, covariantly represented on $L^2(M,d\nu)$ for some compact
manifold $M$.  The automorphisms $\alpha _t$ will be of the form
$\alpha ^H_t(A) = U^*_tAU_t$ where $U_t=\expit H$ for some
positive elliptic pseudodifferential operator of order 1 on $M$.

In all these examples, the GS construction will come down to a
symbol map
$$\sigma :\A\rightarrow C(SB)$$
where $B\subset T^*M\backslash 0$ is a symplectic cone, and where $SB$
is a section of the cone of the form $\{\sigma _H=1\}$.  The
classical limit state $\omega $ will have the form
$$\omega (A) = \int_{SB}\sa d\mu$$
where $d\mu$ is the normalized surface measure on $SB$ induced
by $H$ and by the symplectic volume measure $\Omega^n$, i.e.\ up
to a scalar, $d\mu = dH\lrcorner \Omega^n$.  Normalized will mean that
$\int_{SB} d\mu =1$.  We will refer to $d\mu$ as the Liouville
measure on $SB$.

We will consider the following four algebras:

(A)  $\A = \bar \Psi^\circ (M)$ (scalar pseudodifferential
operators)

(B)  $\A = \bar \Psi^\circ (M,E)$ (matrix pseudodifferential
operators)

(C)  $\A = \bar \A^\circ_\Sigma$ (co-isotropic operators)

(D)  $\A = \bar {\cal T}^\circ_\Sigma$ (Toeplitz operators).

Here, the bar indicates the norm closure of the usual smooth
subalgebras.

Example (A) has been discussed in detail in the articles [CV], [Sn],
[Z.1-3], [Su] and others, and is only included here to illustrate the terminology and notation in a
familiar context.  The algebras $\A^\circ_\Sigma$ and $\cal T^\circ_\Sigma$ are
probably less familiar, but we will have to assume the reader's
familiarity with them: in particular, with their behaviour under
composition with other types of Fourier Integral operators. For
background on $\A^\circ_\Sigma$ we refer to Guillemin--Sternberg
([GS], [G.2]) and for ${\cal T}^\circ_\Sigma$ we refer to Boutet de
Monvel--Guillemin ([B.G], [B]).\medskip

\noindent(A)  $\A = \overline{\Psi^\circ(M)}$.

(a)  $\partial M = \phi$.\smallskip

Let $H\in \Psi^1(M)$ be positive elliptic, and let $H\phi_j =
\lambda _j\phi_j$, $\langle \phi_i, \phi_j\rangle=\delta_{ij}$,
denote its spectral data.  We set
\begin{gather*}\omega _j(A) = \langle A\phi_j, \phi_j\rangle\\
\omega _\lambda  = \frac{1}{N(\lambda )}\sum_{\lambda _{j} \leq
\lambda } \omega _j\\
N(\lambda ) = \#\{j:\lambda _j \leq \lambda \}\\
\omega (A) = \int_{S^{*}M}\sa d\mu\end{gather*}
where $S^*M\subset T^*M$ is the level set $\{H=1\}$.

We observe that $\{\omega _j\}$ are normal invariant ergodic
states of $(\bar\Psi^\circ, \R,\alpha ^H_t)$.  Also, that $\omega $
is a (non-normal) invariant  state by virtue of the Egorov
theorem $\sigma (\alpha ^H_t(A)) =\sa \circ G^t$, where $G^t$ is
the Hamilton flow of $\sigma _H$ on $\{H=1\}$. 
 Condition (c) of
Definition (1.1),
$$\weaklim_{E\rightarrow \infty} \omega _E =\omega $$
is well-known and can be proved by studying the principal term at $t=0$ of the
distribution trace, $\Tr AU_t$.  Indeed,  by the calculus of Fourier Integral
operators $\Tr AU_t$ is a Lagrangean distribution on $\R$, whose
symbol at $t=0$ is, up to universal constants, essentially $\omega
(A)$.  Condition (c) then follows by a Tauberian argument.  For further
details we refer to [HoIV, \S29] or
[G.1]. The other conditions of Definition (1.1) are obvious, and it is 
straightforward that the classical limit system is the geodesic flow on
$S^*M.$  Hence the original system is quantized abelian.

By Theorems 1 and 2,  ergodicity of $G^t$ on $S^*M$ will imply the
quantum ergodicity of $(\bar \Psi^\circ ,\R,\alpha ^H_t)$.  In
particular, there is a subsequence $\{\varphi_{j_{k}}\}$ of
eigenfunctions of density one such that
$$\lim_{k\rightarrow \infty} (A\varphi_{j_{k}}, \varphi_{j_{k}}) =
\omega (A)\;,$$
and $\langle A\rangle = \omega (A) I + K$, where $\|\Pi_\lambda
K\Pi_\lambda \|_{\HS} = o(N(\lambda ))$; here $\Pi_\lambda  =
\sum_{\lambda _{j}\leq \lambda }\varphi_j \otimes
\varphi^*_j$.

(b)  $\partial M$ diffractive.

If $\partial M\neq \phi$, the proof that classical ergodicity (now of the
billiard flow) implies quantum ergodicity becomes more complicated. 
The principal difficulty is that $\alpha _t(A) = U^*_tAU_t$ no longer
necessarily defines an automorphism of the algebra of pseudodifferential
operators on $M$.  In the case of manifolds with diffractive boundary,
Farris' extension of the Egorov Theorem (which is carried out from the
$C^*$ algebra point of view) is sufficient for the proof of
quantum ergodicity using Theorem 1.  For further discussion and a generalization
to manifolds with piecewise smooth boundary and
ergodic billiards, we refer to 
[Z.Zw]. \medskip

\noindent(B)  $\A = \bar \Psi^\circ (M,E)$, where $E\rightarrow M$ is a
real rank $n$ vector bundle.\smallskip

The new feature is that symbols are now matrix valued.  

As above, we let $H\in \Psi^1(M,E)$ be positive elliptic.  Then
$$\sigma _H (x, \xi):(\pi^*E)_{(x,\xi)}\rightarrow
(\pi^*E)_{(x,\xi)}$$
where $\pi:T^*M \rightarrow M$ is the natural projection,
$\pi^*E\rightarrow T^*M$ is the pulled back bundle, and
$(\pi^*E)_{(x,\xi)}$ is its fiber over $(x, \xi) \in T^*M$.  Unless $\sigma
_H(x,\xi) = h(x,\xi)\Id$ for some scalar symbol $h(x, \xi)$, conjugation by
$U_t = \expit H$ will not define an automorphism of $\hat \Psi^\circ
(M,E)$.  This is of course the problem of generalizing Egorov's theorem to
systems; see Cordes [C].

To obtain a $C^*$ dynamical system satisfying the hypothesis of
Theorem 1, we will need to place some conditions on $\sigma _H$
and possibly restrict to a subalgebra of $\Psi^\circ (M,E)$.  The
condition on $\sigma _H$ is that it have constant multiplicities as
$(x,\xi)$ varies over $S^*M$.  Let us consider just the two
extremes:

(i)  $\sigma _H (x, \xi) = h(x,\xi)\Id$ (real scalar type)

(ii)  $\sigma _H(x,\xi)$ has real distinct eigenvalues $\lambda
_1(x,\xi)<\lambda _2(x,\xi)<\cdots <\lambda _m(x,\xi)$ with
$\lambda _{j+1} (x,\xi) - \lambda _j(x,\xi)\geq C (1+|\xi|)$ for
some $C>0$ (strictly hyperbolic type).

In either case, let $\sigma  '_H$ denote the symbolic commutant of
$\sigma _H$, i.e.\ the matrix valued symbols $\sigma (x,\xi)$ on
$T^*M$ such that $[\sigma _H(x,\xi),\sigma (x,\xi)]=0$ for all
$(x,\xi)$.  In case (i), $\sigma '_H$ constants of all
$\End(E)_{(x,\xi)}$-valued symbols.  It follows as in the scalar
case that $\alpha ^H_t$ is an automorphism of the full $\bar
\Psi^\circ (M,E)$ and that $\sigma (\alpha ^H_t(A)) =\sa \circ G^t$
where $G^t$ is the Hamilton flow of $\sigma _H$.  Analysis of $\Tr
AU_t$ at $t=0$ leads as above to the formula
$$\weaklim_{E\rightarrow \infty}\omega _E=\omega $$ 
where $\omega (A) = \int_{S^{*}M}\tr\sa d\mu$.  Ergodicity of
$\omega $ is then equivalent to ergodicity of $G^t$ on $S^*M$.  We have:

\subsection*{(3.1) Corollary}  {\it If $H$ is of real scalar type and
$G^t$ is ergodic, then $(\bar\Psi^\circ (M,E), \R, \alpha ^H_t)$ is
quantum ergodic.}

 As special cases, one could
let $E = \Lambda^kT^*M$, and $H=\sqrt{\Delta_k}$, where $\Delta_k$ is
the Laplacian on $k$-forms.  If $\{\eta_j\}$ is an orthonormal basis of
eigenforms, one obtains $(A\eta_{j_{k}},\eta_{j_{k}})\rightarrow \omega
(A)$ along a density one subsequence.  In particular, $|\eta_{j_{k}} |^2
(x)\overset{\omega }{\rightarrow }1$ where $|\eta|(x)$ is the norm of
$\eta$ at $x$.  Similarly for $H=(\not\partial ^*\not\partial )$ where
$\not\partial $ is the Dirac operator on a spin bundle.  

In the strictly hyperbolic case, $\sigma '_H$ consists of matrix
valued symbols of the form
$$\sigma (x,\xi) = \sum^n_{i=1} a_i (x,\xi) \pi_i
(x,\xi)\;,\leqno(3.2)$$
where $\pi_i (x,\xi)$ is the eigenprojection on $(\pi^*E)_{(x,\xi)}$
corresponding to $\lambda _i(x,\xi)$, and where $a_i (x,\xi)$ is a
scalar symbol of order $0$.

Let $G^t_i:T^*M \rightarrow T^*M$ denote the Hamilton flow of
$\lambda _i$, and let
$$\sigma \circ G^t (x,\xi) = \sum^n_{i=1} a_i (G^t_i(x,\xi))\pi_i
(x,\xi)\;.$$
The map $\sigma \mapsto \sigma \circ G^t$ defines an
automorphism of $\sigma '_H$.  By the Egorov theorem of Cordes
(loc.\ cit.) for each $\sigma \in \sigma '_H$ one can construct an
operator $A\in \Psi^\circ (M,E)$ such that
\begin{gather*}
\sa=\sigma \\
\alpha ^H_t (A) \in \Psi^\circ (M,E)\tag{3.3}\\
\sigma (\alpha ^H_t(A)) = \sa \circ G^t\;.\end{gather*}
Let us define $\Psi^\circ _H(M,E) \subset \Psi^\circ (M,E)$ by:
$$\Psi^\circ_H(M,E)=\{A:A \text{ satisfies (3.3)}\}\;.$$
Then $\Psi^\circ_H$ is a non-trivial $*$-subalgebra, and hence
$(\bar\Psi^\circ_H, \R, \alpha ^H_t)$ is a $C^*$ dynamical system. 
Analysis of $\Tr AU_t$ leads to the limit formula:
$$\lim_{E\rightarrow \infty} \omega _E (A) = \omega
(A):=\sum^n_{j=1}\int_{\{\lambda _j = 1\}}a_j d\mu_j\quad (A\in
\Psi^\circ _H)$$
where $d\mu_j$ is the Liouville measure on $\{\lambda _j = 1\}$. 
Ergodicity of $\omega $ is equivalent to the ergodicity of each of
the flows $G^t_j$.  Therefore we have:

\subsection*{(3.4) Corollary} $(\bar \Psi^\circ_H, \R, \alpha ^H_t)$ {\it
is quantum ergodic if $H$ is of strictly hyperbolic type and if all the
flows $G^t_j$ are ergodic.}\medskip

\noindent(C)  $\A=\bar\A^\circ_\Sigma = \overline{\Pi R^\circ_\Sigma
\Pi}$ ( a corner of a co-isotropic or flowout algebra).\smallskip

Here the algebra is (the $C^*$ closure of) a ``corner" of the $*$
algebra $R_\Sigma $ associated to a co-isotropic cone $\Sigma \subset
T^* M\backslash O$, in the sense of Guillemin--Sternberg
([G.S], [G.2]).  These algebras arise in the reduction of
quantum systems with symmetries, and have already been
studied in connection with quantum ergodicity in
([Z.4], [S.T]).  We briefly review the definition and properties of
$R_\Sigma $.  For more details, we refer to [G.S].

Let ${\cal N}$ denote the null foliation of $\Sigma $.  The equivalence
relation in $\Sigma \times\Sigma $ of belonging to the same leaf of
${\cal N}$ is a Lagrangean relation $\Gamma\subset T^*(M\times
M\backslash O)$ such that $\Gamma\circ \Gamma = \Gamma$ and
$\Gamma^t = \Gamma$.  Hence the algebra $R_\Sigma :=I^*(M\times
M,\Gamma)$ of Fourier Integral operators associated to $\Gamma$
is a $*$ algebra.  We will assume that the foliation ${\cal N}$
defines a fibre bundle over a base $B$, necessarily a symplectic
cone.  The symbol algebra $S_\Sigma $ corresponding to $R_\Sigma $
can then be identified with functions $T(b,\cdot,\cdot)$ on $B$
with values in the smoothing operators on the fibres $F_b$ of
$\Sigma \rightarrow B$.

We will assume further that the fibers $F_b$ are compact.  Then
(the weak closures of) $S_\Sigma $ and $R_\Sigma $ are factors of type
$I_\infty$, i.e.\ contain minimal projections.  If $\Pi\in R_\Sigma $
is such a projection, then its symbol $\sigma (\Pi)$ is for each $b$
a rank one projection $\sigma (\Pi) (b,\cdot,\cdot)$ on $L^2(F_b)$.

Let us fix one minimal projection $\Pi\in R_\Sigma $ and consider
the corner $\Pi R_\Sigma \Pi$ of $R_\Sigma $.  By [G.S],
[G.2] each element of $\Pi R_\Sigma \Pi$ can be presented in
the form $\Pi A\Pi$, where $A\in \Psi^*_\Pi = \{ A\in
\Psi^*:[A,\Pi]=0\}$.  If $A\in\Psi^*_\Pi$, its symbol $\sa$ is
constant on the fibers $F_b$ and can be identified with a function
$\hat{\sa}$ on $B$.  Also, $\sigma (\Pi A\Pi) =
\hat{\sa}\sigma_\Pi$; so that the symbol algebra of $\Pi R^\circ_
\Sigma \Pi$ can be identified with homogeneous functions on $B$ of
order $0$.

Now let $H\in \Psi^1_\Pi$ be positive elliptic.  Then $\alpha ^H_t$
defines an automorphism of $\Pi R^\circ_\Sigma  \Pi$ for $t\in
\R$.  The faithful covariant representation in this example is on the
Hilbert space $\Hh_\Pi = \range (\Pi)$, which is heuristically the
quantization of the symplectic quotient $B$.  Since $H\in
\Psi^1_H$, $U_t=\expit H$ operates on $\Hh_\Pi$ with discrete
spectrum.  Let $\{\varphi^\Pi_j\}$ denote an orthonormal basis of
$\Hh_\Pi$ of eigenvectors of $H$, and let $\{\omega ^\Pi_j\}$
denote the corresponding invariant states.  The trace $\Tr \Pi A
U_t$ can be analyzed as a  composition of Fourier Integral
operators, which leads to the limit formula
$$\lim_{E\rightarrow \infty} \omega ^\Pi_E(A) = \int_{\SB} \hat
\sa d\mu\quad \quad(A\in \Psi^\circ _\Pi)$$
where $\omega ^\Pi_E$ is the microcanonical ensemble for
$\Hh_\Pi$.  In other words, if $\Pi_E$ denotes the full spectral
projection for $H$ on the interval $[0,E]$, then 
$$\omega ^\Pi_E = \frac{1}{\Tr \Pi \cdot \Pi_E} \sum_{\lambda _j
\leq E} \omega ^\Pi_j\;.$$
Ergodicity of the state $\omega_\Pi(A) = \int_{\SB} \hat
\sa d\mu$ is equivalent to ergodicity of the quotient flow $G^t$ on
$(SB,\mu)$.  Hence we have: 

\subsection*{(3.5) Corollary} $(\Pi\bar{R}^0_\Sigma \Pi, \R, \alpha^H_t) $
{\it is quantum ergodic if the quotient flow $G^t$ on $\SB$ is ergodic.}

It would be interesting to study the non-fibrating case where the
fibers are not compact.\medskip

\noindent(D) $\A = \bar{\cal T}_\Sigma $ (a Toeplitz algebra)
\smallskip

Here, $\Sigma\subset T^* M\backslash 0$ is a closed symplectic cone.  By
[B.G] it has a Toeplitz structure; that is, there is an associated
projector  $\Pi_\Sigma$ on $L^2(M)$ with the microlocal properties of
the  Szeg\"o projector of a strictly pseudo convex domain $\Omega$.  
In this case, for instance, $\partial\Omega$ has a natural contact
structure $\alpha$, $\Sigma = \{(x, rdx) : x \in \partial
\Omega, r>0\} \subset T^* (\partial\Omega)\backslash 0$ and
$\Pi_\Sigma $ is the orthogonal projection $L^2(\partial
\Omega)\rightarrow H^2(\partial \Omega)$ onto boundary values of
holomorphic functions on $\partial \Omega$ which lie in $L^2(\partial
\Omega)$.  In general, the range of $\pis$ is a Hilbert space
$\Hh_\Sigma $ which one thinks of as the quantization of $\Sigma$. 

The Toeplitz algebra $\T_\Sigma $ is the algebra of elements $\pis
A\pis$ with $A\in \Psi^*(M)$.  One can again represent each
element in the above form with $A\in \Psi^*_{\pis} = \{B\in
\Psi^*:[B,\pis]=0\}$ ([B.G,Proposition 2.13 and p.\ 82]).  Hence
$\T_\Sigma  \simeq \Pi \Psi^*_\Pi \Pi/\vartheta(\Pi)$ where
$\vartheta_\Pi = \{A\in \Psi^*_\Pi:A\Pi = 0\}$.  Here (and
henceforth) we write $\Pi$ for $\pis$. $\T_\Sigma $ has a faithful
covariant representation on $\Hh_\Sigma $.

The principal symbol $\sigma (\Pi A\Pi)$ of an element of
$\T_\Sigma $ may be identified with $\sigma  |_\Sigma$, and the symbol
algebra for $\bar \T^\circ_\Sigma $ with the algebra of continuous
homogeneous functions of degree $0$ on $\Sigma$.  (See [B.G].)

Let $H\in \Psi'_\Pi$ be positive elliptic.  Then $U_t = \expit
H|_{\Hh_{\Sigma }}$ defines a unitary representation of $\R$ with
discrete spectrum.  As before we let $\{\phi_j\}$ be an
orthonormal basis of eigenfunctions of $H$ in $\Hh_\Sigma$, let
$\omega _j$ be the corresponding states, let  $\Pi_E$ project to span
$\{\phi_j:\lambda _j\leq E\}$ and let $$\omega_E =\frac{1}{\rank
(\Pi_E\Pi)}\sum^{\rank \Pi_{E}\Pi}_{\lambda _j\leq \lambda } \omega
_j\;.$$ Analysis of the trace $\Tr \Pi U_t A\quad (A \in \Psi^\circ_\Pi)$
shows that $\omega _E\overset{\weak^*}{\rightarrow }\omega
_\Pi $ with
$$\omega _\Pi (\Pi A\Pi) = \int_{S\Sigma } \sa d\mu$$
where $S\Sigma  = \{\sigma _H = 1\} \cap \Sigma $.  The composition
theorem for Fourier Integral and Hermite operators [B.G.,\S7]
shows that 
$$\sigma (\alpha ^H_t(\Pi A\Pi)) = \sa \circ G^t|_\Sigma $$
where $G^t$ is the Hamilton flow of $\sigma _H$ on $\Sigma $. 
Ergodicity of $\omega _\Pi$ is equivalent to ergodicity of the
sub-flow $G^t|_\Sigma $ with respect to $\mu$.  Hence, 

\subsection*{(3.6) Corollary} {\it $(\bar\T^\circ_\Sigma ,\R,\sigma
^H_t)$ is quantum ergodic if $G^t$ is ergodic on $(\Sigma ,\mu)$.}

Let us note that if $[\Pi,H]=0$, then $\Pi U_t\Pi = \Pi\expit(\Pi
H\Pi)\Pi$.  Hence we may view the generator of the covariant
representation of $\R$ as the Toeplitz operator $\Pi H\Pi$.  

\subsection*{(3.7) Example}  Suppose  that $H_1$ is positive elliptic,
for instance $H_1 = \sqrt{\Delta}$ for some Riemannian metric, and let
$\gamma$ be a closed orbit for the Hamilton flow of $H_1$ on $S^*M$. 
Then the cone $\Sigma  =\R^+\gamma$ through $\gamma$ is a symplectic
submanifold of $T^*M\backslash 0$.  Let $\Pi_\Sigma $ be a Toeplitz
structure for $\Sigma $.  Ideally we  would like $[H_1, \Pi_\Sigma ]=0$
but it is not generally possible to construct $\pis$ with this
property unless the whole geodesic flow $G^t_1$ of $H_1$ is
periodic [B.G.,Appendix].  However, by [B.G.,Proposition
2.13] for any choice of $\pis$ we can find $H\in \Psi^1$, such
that $[H,\pis]=0$, $\sigma _H|_{\Sigma } = \sigma _{H_{1}}|_\Sigma $,
and $\pis H_1\pis = \pis H\pis$.  Since $\sigma _{H_{1}}$ and
$\sigma _H$ generate the same Hamilton flows on $\Sigma $,
$\gamma$ is a periodic orbit of the flow of $\sigma _H$. 
Obviously, the uniform measure $\mu_\gamma$ is ergodic for this
flow.  It follows from (3.6) that $(\bar \T^\circ_\Sigma , \R, \alpha
^H_t)$ is quantum ergodic, hence the eigenfunctions
$\phi^\gamma_j$ of $\pis H_1\pis$ concentrate on $\gamma$ in
the limit $j\rightarrow \infty$.

Note that the $\phi^\gamma _j$ are actual eigenfunctions of $H_1$ in
$\Hh_\Sigma$.  Since $H_1$ and $H$ are close in a microlocal
neighborhood of $\gamma $, the $\{\phi^\gamma_j\}$ may be
viewed as a kind of quasi-mode for $H$ associated to $\gamma$. 
The approximation here is very weak, of course; the
$\phi^\gamma_j$ concentrate (or ``scar") along $\gamma$, while
it is  doubtful that any sequence of $H$-eigenfunctions has
this property (see however [H]).

\addtolength{\baselineskip}{-4pt}

\end{document}